\documentclass[aps,prl,reprint,superscriptaddress]{revtex4-2}
\usepackage{graphicx}% Include figure files
\usepackage{dcolumn}% Align table columns on decimal point
\usepackage{bm}% bold math
\usepackage{amsmath}
\usepackage{lineno}
\usepackage{url}
\usepackage{bookmark}
\usepackage{natbib}
\usepackage{color}
\usepackage{hyperref}
\hypersetup{
    colorlinks=true,
    linkcolor=blue,
    filecolor=blue,      
    urlcolor=blue,
    citecolor=blue,
    }

\begin{document}
%\linenumbers

\title{\textbf{Intense X-ray Vortices Generation via Wavefront Shaping in High-gain Free-Electron Lasers}}

\author{Zhikai Zhou}
\affiliation{ShanghaiTech University, Shanghai 201210, China}
\author{Yin Kang}
\affiliation{Shanghai Institute of Applied Physics, Chinese Academy of Sciences, Shanghai 201800, China}
\affiliation{University of Chinese Academy of Sciences, Beijing 100049, China}
\author{Weishi Wan}
\affiliation{ShanghaiTech University, Shanghai 201210, China}
\author{Chao Feng}
\email[]{fengc@sari.ac.cn}
\affiliation{Shanghai Advanced Research Institute, Chinese Academy of Sciences, Shanghai 201210, China}

%\date{\today}

\begin{abstract}
    The x-ray vortex optical beam, distinguished by its topological charge and orbital angular momentum, offers new insights in probing complex electronic structures, enhancing material characterization, and advancing high-resolution imaging techniques. Here we propose a novel and reliable method to generate intense x-ray vortices with tunable wavelengths in high-gain free-electron lasers (FELs). By simply adjusting the wavefront tilt of the radiation pulse during the FEL gain process, high-quality vortex beam can be amplified until saturation. Compared with existing methods for FEL vortex generation, the proposed technique offers broader wavelength tunability and can be easily implemented in high-gain FEL facilities, regardless of the operating modes.
\end{abstract}

\maketitle

The rapid development of vortex optical beams carrying orbital angular momentum (OAM) \cite{PhysRevA.45.8185,RN1} in conventional laser fields has led to numerous significant applications, including optical tweezers \cite{doi:10.1126/science.1058591, doi:10.1126/science.1069571,RN2, Shen:12}, super-resolution imaging \cite{RN3,Furhapter:05, PhysRevLett.97.163903,RN4}, and quantum information processing \cite{PhysRevA.88.032305}. These beams are distinguished by their unique helical phase structure, characterized by a phase singularity and an azimuthal phase dependence $e^{-il\phi}$, where $l$ represents the topological charge, indicating the number of $2\pi$ phase rotations around the beam's azimuthal axis within one wavelength, and $\phi$ is the azimuthal angle. The intense potential of this specific property to transform various fields, from fundamental physics to material science and biology, has generated considerable interest \cite{PhysRevLett.79.2450, doi:10.1126/science.1100603, RN5, RN6}. Extending OAM technology to the x-ray regime or even higher photon energies is particularly attractive, as it could provide unprecedented insights into nanoscale structures, dynamics, and nuclear physics \cite{PhysRevB.87.121201, photonics4020028, np-photoelectric, Wang:23, McCarter_2024, IVANOV2022103987}. However, conventional laser-based methods are often limited by their inability to achieve the short wavelengths required for x-ray production.

Modern free-electron lasers (FELs) have emerged as one of the most promising avenues for generating coherent x-ray sources with high intensity. The advent of FELs has represented a paradigm shift in x-ray science, providing ultra-high brightness, femtosecond to attosecond pulse durations, and tunability across a broad spectral range \cite{RN8, RN9, RN7, article, RN10, RN12, RN13, Feng:22, RN19, RN20}. These attributes make FELs particularly suitable for investigating x-ray OAM. One pioneering approach to generate OAM in FELs focuses on using helical undulators, which naturally produce higher harmonics carrying OAM because of the helical trajectory of electrons within the undulator's magnetic field \cite{PhysRevLett.102.174801}. However, this method often results in relatively weak OAM signal due to the high harmonic generation process in the undulator. To address this issue, a more elaborate technique has been developed, involving the interaction of a seed laser with an electron beam in a helical undulator, allowing the generation of vortex FEL beams at the fundamental frequency \cite{PhysRevLett.106.164803}. Additionally, clever methods of using seed lasers with tailored transverse phases to shape electron bunches into helical patterns, or using spiral zone plates to manipulate the transverse phase of the radiaiton pulse, thus achieving EUV and x-ray OAM light, have been explored \cite{PhysRevLett.109.224801, PhysRevLett.112.203602, PhysRevX.7.031036, Sun_Wang_Feng_Tu_Fan_Liu_2021, yan2023self}. Although these methods show promise, they are often limited by the availability and quality of seed lasers and optical elements (such as different materials for spiral zone plates across various wavelength ranges), and they often lack broad wavelength tunability. Furthermore, achieving precise control over the electron beam's microstructure and maintaining the dominance of the OAM mode throughout the FEL amplification process remains a significant technical challenge. 

In this letter, we propose a novel and simple method for generating intense x-ray vortex beams via mode conversion in high-gain FELs. Our approach takes advantage of the natural amplification process of FELs to produce high-quality OAM beams with tunable wavelengths. This technique offers several advantages over existing methods, such as broader wavelength coverage and ease of implementation in current high-gain FEL facilities. 

\begin{figure*} %%% Fig 1
    \begin{minipage}{0.75\linewidth}
    \centering
    \includegraphics[width=\linewidth]{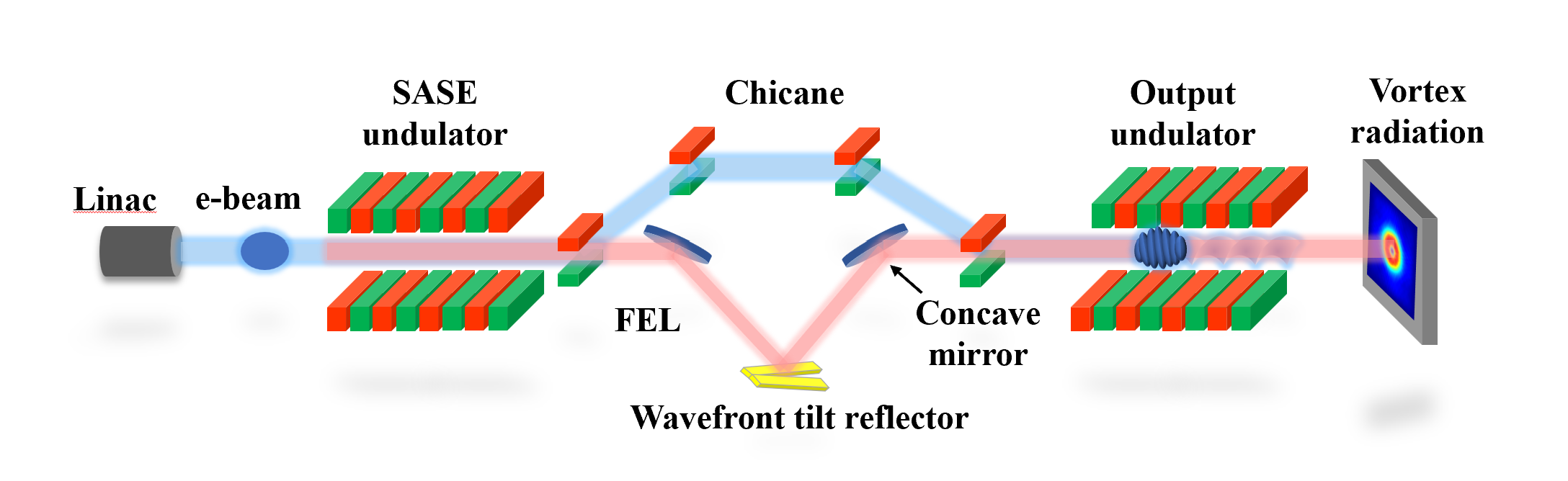}
    \put(-360,100){(a)}
    \end{minipage}
    \begin{minipage}{0.85\linewidth}
    \centering
    \includegraphics[width=\linewidth]{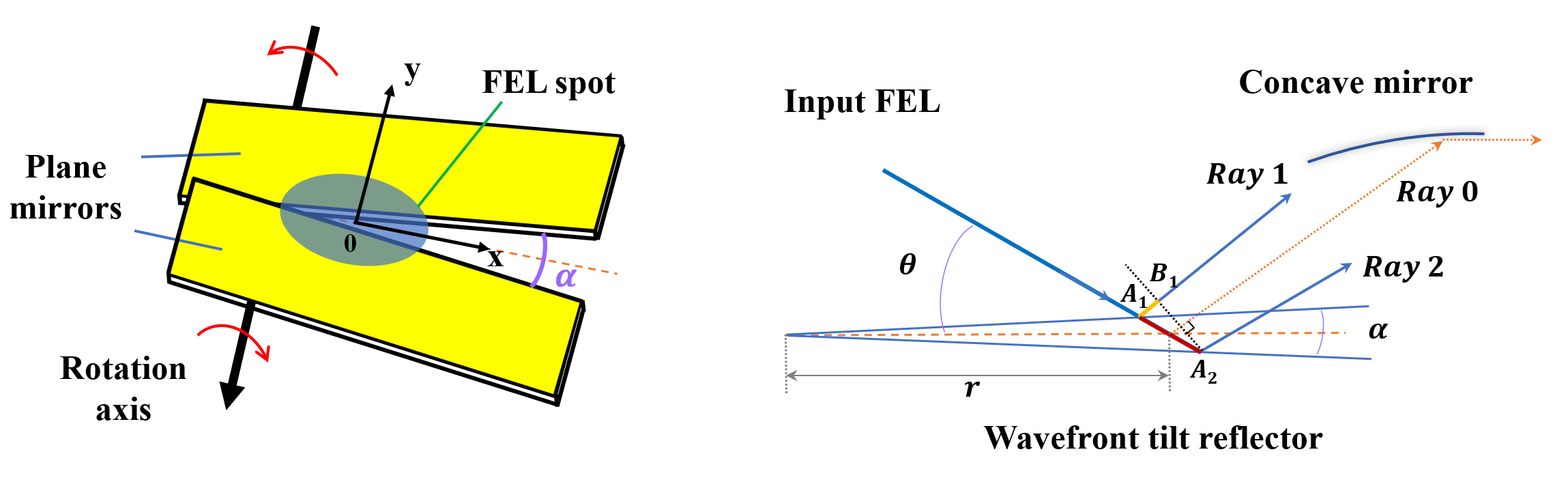}
    \put(-380,130){(b)}
    \put(-210,130){(c)}
    \end{minipage}
    \caption{Diagram for the proposed method. (a) Schematic layout for x-ray OAM generation based on SASE. (b) The structure of the wavefront tilt reflector. (c) Two-dimensional geometry of the optical propagation of the FEL pulse in the chicane.}
    \label{Fig1}
\end{figure*}

The key to the proposed method is subtly adjusting the wavefront of the radiation pulse during the FEL amplification process, leveraging diffraction effects and the FEL gain to continually twist the radiation profile into a vortex beam at saturation. The schematic layout of the proposed technique is shown in Fig. \ref{Fig1}, where the self-amplified spontaneous emission (SASE) mode \cite{Kondratenko1980GENERATINGOC, BONIFACIO1984373} is adopted as an example to demonstrate the physical process. Similarly to the self-seeding technique \cite{soft-x-ray-self-seeding-principle, soft-x-ray-self-seeding-experiment, hard-x-ray-self-seeding-principle, hard-x-ray-self-seeding-experiemnt}, the undulator is separated into two parts by a bypass chicane, as shown in Fig. \ref{Fig1}(a). After sufficient amplification in the SASE undulator, the FEL pulse and the electron beam are separated in the chicane, and the transverse phase of the radiation pulse is shaped through the optical elements. The electron beam timing delay induced by the chicane should match the optical delay caused by these optical elements, ensuring that the radiation pulse serves as a seed in the output undulator.

There are three optical elements in the chicane: a concave mirror to reflect the input FEL beam and adjust its position, a wavefront tilt reflector to slightly tune the wavefront tilt to shape the transverse pattern of the FEL beam, and another concave mirror to focus the output FEL beam and direct it into the following undulator. The wavefront tilt reflector consists of two plane mirrors that can be independently rotated around the same axis. The angle between the two mirrors is indicated as $\alpha$, as shown in Fig. \ref{Fig1}(b). We assume that $\alpha$ is very small ($\alpha \ll 1$) and negligible compared to the incidence angle $\theta$ of the FEL beam. The input FEL pulse should be centrally placed between these two plane mirrors, creating different wavefront tilts in the two halves of the FEL spot after reflection, as shown in Fig. \ref{Fig1}(c). It is also possible to use spiral phase mirrors \cite{Campbell:12, 10.1073/pnas.1616889113} or computer-generated holograms \cite{Terhalle:11} to convert input FELs into vortex beams. However, using two-plane mirrors offers the advantages of easier fabrication, simpler implementation, and better flexibility in wavelength tuning.

The transverse phase of the output FEL pulse will be shaped by the optical path difference induced during reflection from the two tilted mirrors, as illustrated in Fig. \ref{Fig1}(c). For a point located at a radial distance $r$ (along the axis x) from the rotation axis, an incidence ray is reflected by the upper and lower mirrors at points $A_1$ and $A_2$, respectively, splitting into to two output rays: ray 1 (from the upper mirror) and ray 2 (from the lower mirror). Although the non-zero tilt angle $\alpha$ introduces a slight non-parallelism between the rays, this effect is negligible under the small-angle approximation $\alpha \ll 1$. To quantify the optical path difference $\Delta s$, we define an auxiliary point $B_1$ such that the segment $A_2B_1$ is orthogonal to the propagation direction of ray 1. The optical path difference between the two rays is then geometrically determined by the difference in path lengths: $\lvert A_1A_2\rvert-\lvert A_1B_1\rvert$, and can be represented as
\begin{equation}
    \Delta s={2r\alpha\sin{\theta}}.
    \label{Eq1}
\end{equation}
Consequently, the relative variation of the transverse phase $\Delta\varphi$ can be expressed as $\Delta\varphi=2\pi\Delta s/\lambda$, where $\lambda$ is the central wavelength of the FEL, and we get the transverse phase difference at the radial distance $r$:
\begin{equation}
    \Delta\varphi={4\pi r\alpha\sin{\theta}/\lambda}.
    \label{Eq2}
\end{equation}
 This phase modulation is critical for shaping the wavefront of the output pulse. The interaction region, defined as the spatial overlap between the FEL beam and the electron beam in the downstream undulator, must be optimized for FEL gain. To ensure full overlap, the FEL beam size should be slightly larger than the electron beam size, with the latter determining the effective interaction region. For a given optical spot projection size $\sigma$ of the FEL beam along the dividing line of the two mirrors, a portion with a diameter of $d$ will interact with the electron beam. For generating optical vortices with a topological charge of $\pm1$, the azimuthal phase variation must satisfy $\pm2\pi$ within a wavelength $\lambda$. Substituting $\Delta\varphi=2\pi$ into Eq. (\ref{Eq2}) yields the required tilt angle:
\begin{equation}
    \alpha = {\frac{\lambda}{2d\sin\theta}}. 
    \label{Eq3}
\end{equation}
For a central wavelength of 5 nm, incidence angle $\theta$ of 175 mrad, and interaction diameter $d$ of 100 $\mu$m, the calculated $\alpha$ is about 144 $\mu$rad.

The x-ray transport process through the concave mirror can be studied using the wave optical propagation method \cite{Goodman1969IntroductionTF, PhysRevSTAB.18.030708}. After the concave mirror with a focal length of $f$, the complex optical field of the radiation $\widetilde{E}(x_i,y_i)$ will be changed to
\begin{equation}
    \widetilde{E}'(x_i,y_i)=\widetilde{E}(x_i,y_i)e^{-\frac{ik}{2f}({x_i}^2 + {y_i}^2)},
    \label{Eq4}
\end{equation}
where $x_i$ and $y_i$ denotes the initial transverse position of the field before the concave mirror, $k$ is the wavenumber of the FEL pulse. The optical field after a propagating distance $f$ can be derived from Fresnel optics as
\begin{equation}
    E(x,y)=\frac{e^{ikf}}{i\lambda f}e^{\frac{ik}{2f}({x}^2 + {y}^2)} \times \mathcal{F}[\widetilde{E}'(x_i,y_i)e^{\frac{ik}{2f}({x_i}^2 + {y_i}^2)}],
    \label{Eq5}
\end{equation}
and it can be mathematically simplified to
\begin{equation}
    E(x,y)=\frac{e^{ikf}}{i\lambda f}e^{\frac{ik}{2f}({x}^2 + {y}^2)} \times \mathcal{F}[\widetilde{E}(x_i,y_i)].
    \label{Eq6}
\end{equation}

\begin{figure} %%% Fig 2
    \centering
    \includegraphics[width=\linewidth]{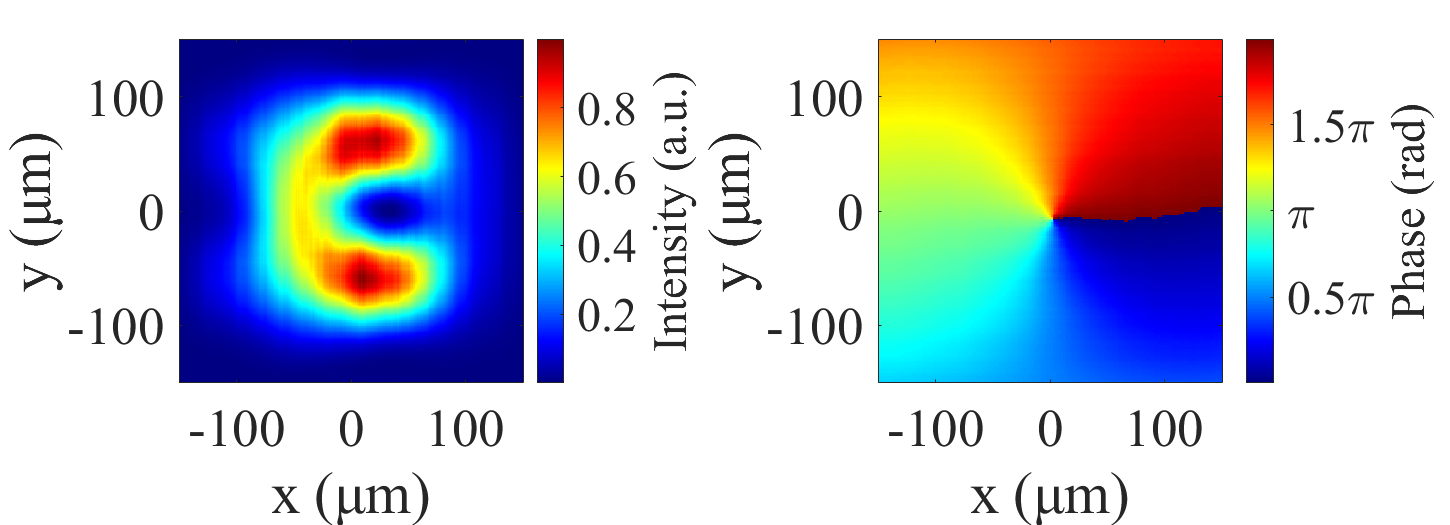}
    \put(-195,90){(a)}
    \put(-80,90){(b)}
    \caption{The intensity (a) and the phase (b) distributions of the output FEL after the chicane.}
    \label{Fig2}
\end{figure}

 Eq. (\ref{Eq6}) indicates that the optical field in the focal plane is the Fourier transformation of the initial field, thus preserving its phase structure. With a focal length of approximately 3 m for the second concave mirror, the transverse intensity and phase of the seeding pulse in the focal plane are depicted in Fig. \ref{Fig2}(a) and (b), respectively. As shown in Fig. \ref{Fig2}(b), the upper and lower halves exhibit oppositely directed wavefront tilts, but the dividing line is no longer clear due to focusing and diffraction during the filed propagation process. The phase difference $\Delta\varphi$ induced by these two mirrors is approximately $\pi$ in the center of the x-ray spot, resulting in a hole in the radiation intensity distribution, as illustrated in Fig. \ref{Fig2}(a). Analysis of the OAM modes of this seeding beam reveals that the component with $l=1$ constitutes 50\% of the total power. Due to the small diffraction angle of the x-ray beam, the transverse distribution of the seeding beam generally retains this pattern at the beginning of the subsequent output undulator.

To demonstrate the feasibility of this method for initiating coherent x-ray OAM radiation, three-dimensional simulations using GENESIS \cite{REICHE1999243} were performed, utilizing typical parameters for a soft X-ray FEL facility, as summarized in Table \ref{Tab1}. The SASE undulator contains five undulators with a length of 3.6 m each and drift sections with a length of 1 m between them. The average $\beta$ function in the undulator is about 10 m. The FEL peak power exceeds 3 GW at the end of the SASE undulator. Subsequently, both the electron beam and the SASE beam were sent into the chicane. The propagation of the SASE beam was simulated using the wave optical propagation method with the same optical parameters as described above. Here we assume a power reflectivity of $80\%$ for each mirror. The shaped radiation beam was then sent into the following undulator to interact with the same electron beam. The microbunching in the electron beam, formed in the previous SASE undulator, was dispersed after passing through the chicane.

\begin{table}[ht]
    \caption{\label{tab:table1}Main parameters used in simulations.}
    \begin{ruledtabular}
        \begin{tabular}{lcc}
            \textrm{Parameters}&
            \textrm{Value}\\
            \colrule
            Electron beam energy & 2.5 GeV\\
            Energy spread & 250 keV\\
            Peak current & 3000 A\\
            Bunch length & 80 fs\\
            Emittance & 0.5 mm$\cdot$mrad\\
            Undulator period & 4 cm\\
            Radiation wavelength & 5 nm
        \end{tabular}
    \end{ruledtabular}
    \label{Tab1}
\end{table}

\begin{figure}[htbp] %%% Fig 3
    \begin{minipage}{0.9\linewidth}
    \centering
    \includegraphics[width=\linewidth]{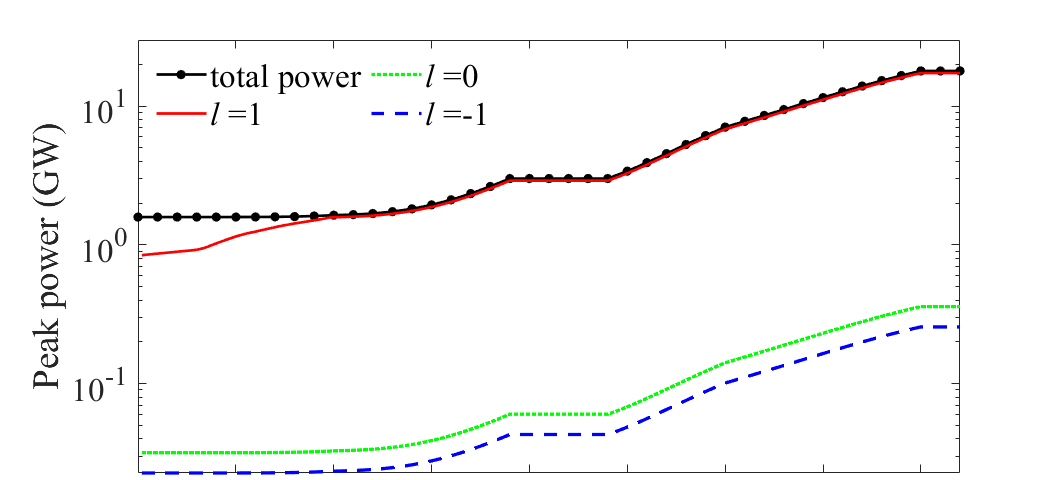}
    \put(-230,90){(a)}
    \end{minipage}
    \begin{minipage}{0.9\linewidth}
    \centering
    \includegraphics[width=\linewidth]{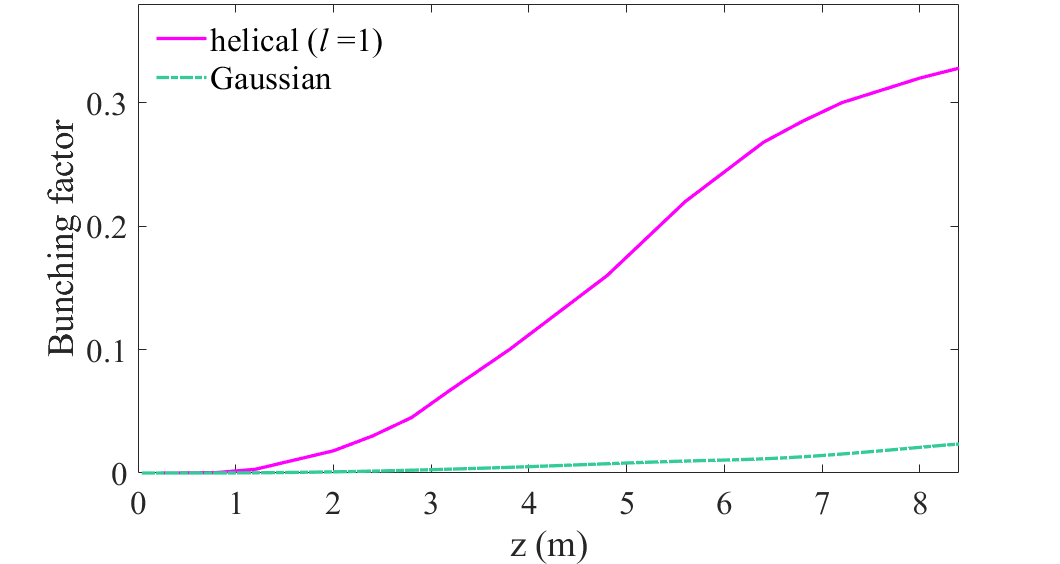}
    \put(-230,115){(b)}
    \end{minipage}
    \caption{Gain curves (a) and evolutions of bunching factors (b) for different FEL modes along the output undulator.}
    \label{Fig3}
\end{figure}

 The output undulator consists of two undulators separated by a drift section with parameters identical to those of the SASE undulators. The focal point of the concave mirror is aligned with the start of the output undulator. In the first 2 meters of the undulator, the seeding beam continuously interacts with the electron beam, inducing microbunchings within it. The wavefront of the microbunching is shaped by the radiation, which in turn modifies the wavefront of the radiation itself. The evolution of the radiation profile and the microbunching is primarily governed by the diffraction of the field and the amplification of the $l=1$ mode. This dynamic interaction continues until the helical microbunching is established in the electron beam. During this process, no net power growth is observed, as illustrated in Fig. \ref{Fig3}(a).

\begin{figure}[ht]%%% Fig 4
    \centering
    \includegraphics[width=\linewidth]{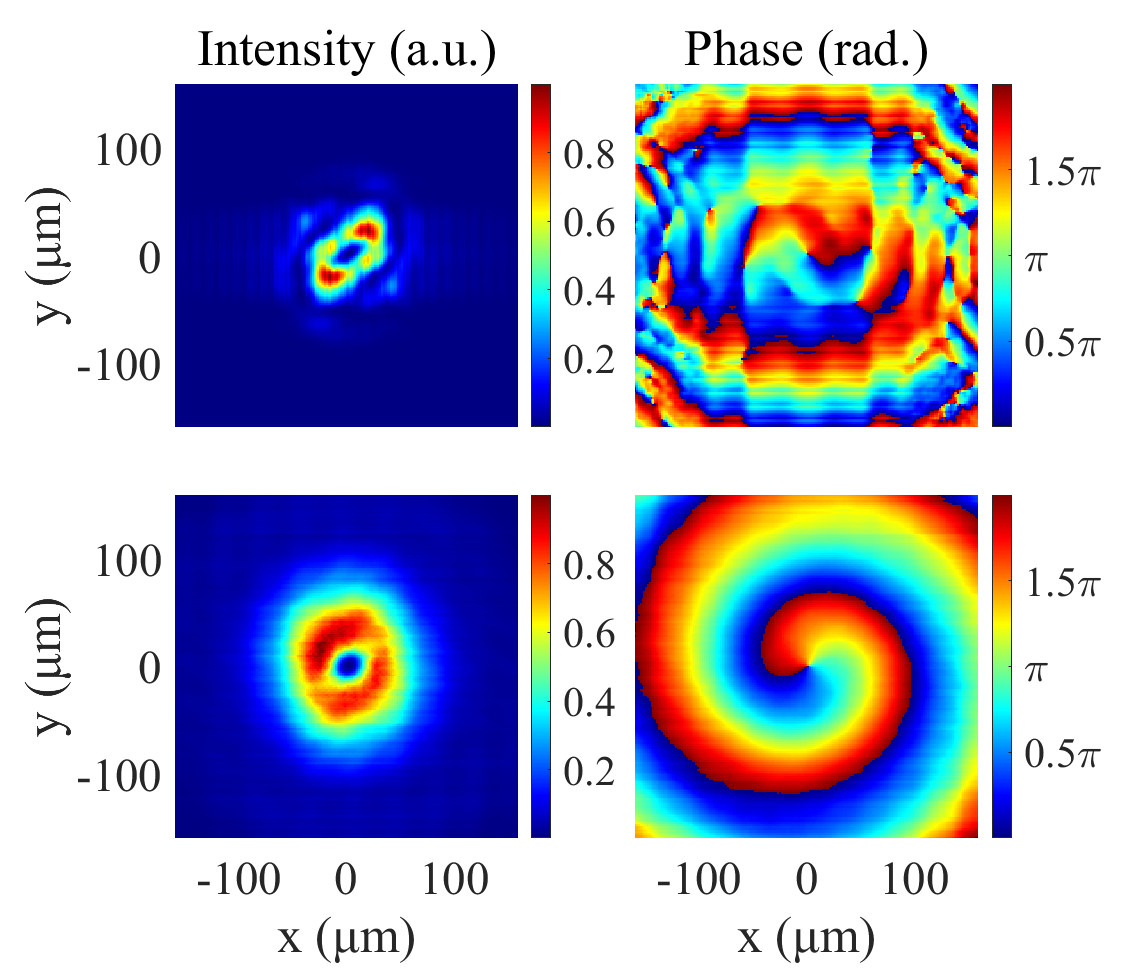}
    \caption{Intensity and phase profiles after 2 m of the output undulator (upper row), and at the exit of output undulator (lower row).}
    \label{Fig4}
\end{figure}

Microbunching can be quantified by the bunching factor \cite{PhysRevLett.109.224801}, $b^{l}(k)=\vert\langle{e}^{-iks-il\phi}\rangle \vert$, where the bracket denotes the average over all coordinates, and $s$ is the longitudinal position along the electron beam. As shown in Fig. \ref{Fig3}(b), the average helical bunching factor across the electron bunch increases from 0 to approximately 2\% within the first 2 meters of the output undulator. As a result, the intensity and phase of the radiation field exhibit a helical twist, as illustrated in the first row of Fig. \ref{Fig4}. Subsequently, the FEL enters the exponential gain regime, where the selection of the transverse mode favors lower order modes, such as the Gaussian mode ($l=0$) and the OAM modes with $l=\pm1$. 

The predominance of the $l=1$ mode in this case arises from gain competition within the output undulator. The seeding beam, shaped by the tilted mirrors, introduces a transverse phase variation ranging from 0 to $2\pi$, aligning with the structure of the mode $l=1$ and promoting its amplification over other modes. The transverse patterns of $l=0$, $l=-1$ and other higher modes developed in the SASE undulator are disrupted by the wavefront tilt reflector, preventing these modes from entering the normal linear gain regime and thereby significantly suppressing their amplification in the output undulator. Consequently, exponential growth culminates in the dominance of the guided OAM mode with $l=1$ at saturation, as shown in Fig. \ref{Fig3}(a). At the exit of the output undulator, the FEL peak power reaches tens of gigawatts, with the $l=1$ mode comprising 97\% of the total power, while the Gaussian mode contributes only 2\%. The characteristic hollow intensity distribution and the helical phase of the vortex radiation at saturation are illustrated in the second row of Fig. \ref{Fig4}. 

The underlying physics of the proposed method involves inducing wavefront tilts in opposite directions on either halve of the FEL spot to facilitate the rotation of the radiation phase front. In the above design, the angle of the wavefront tilt reflector was determined according to Eq. (\ref{Eq3}). Deviations from the ideal value will not affect the trend of vortex formation but will shift the position of the optical singularity, thereby impacting the transverse symmetry of the output vortex. Consequently, it is essential to analyze the adjustment precision of the wavefront tilt reflector. Multiple simulations have been performed with varying tilt angles of the reflector. The main results are summarized in Fig. 5. We found that deviations from the ideal value should be controlled within 10\%, approximately 14.4 $\mu$rad in this context, to ensure optimal vortex quality. Exceeding this threshold will compromise the hollow intensity distribution and phase gradient of the optical vortex. To achieve shorter wavelength vortex beams, stricter requirements for wavefront tilt adjustment precision will be necessary, necessitating careful design of the mechanical system of the x-ray transport line. High-precision mechanical structures  developed for nanoimaging at synchrotron radiation facilities \cite{Mechanical-system-SR, Mechanical-system-CXFEL}, with typical angular positioning resolution and stability ranging from 10 to 30 nrad, can meet the requirements of the proposed technique.

\begin{figure} [ht]%%% Fig 5
    \centering
    \includegraphics[width=\linewidth]{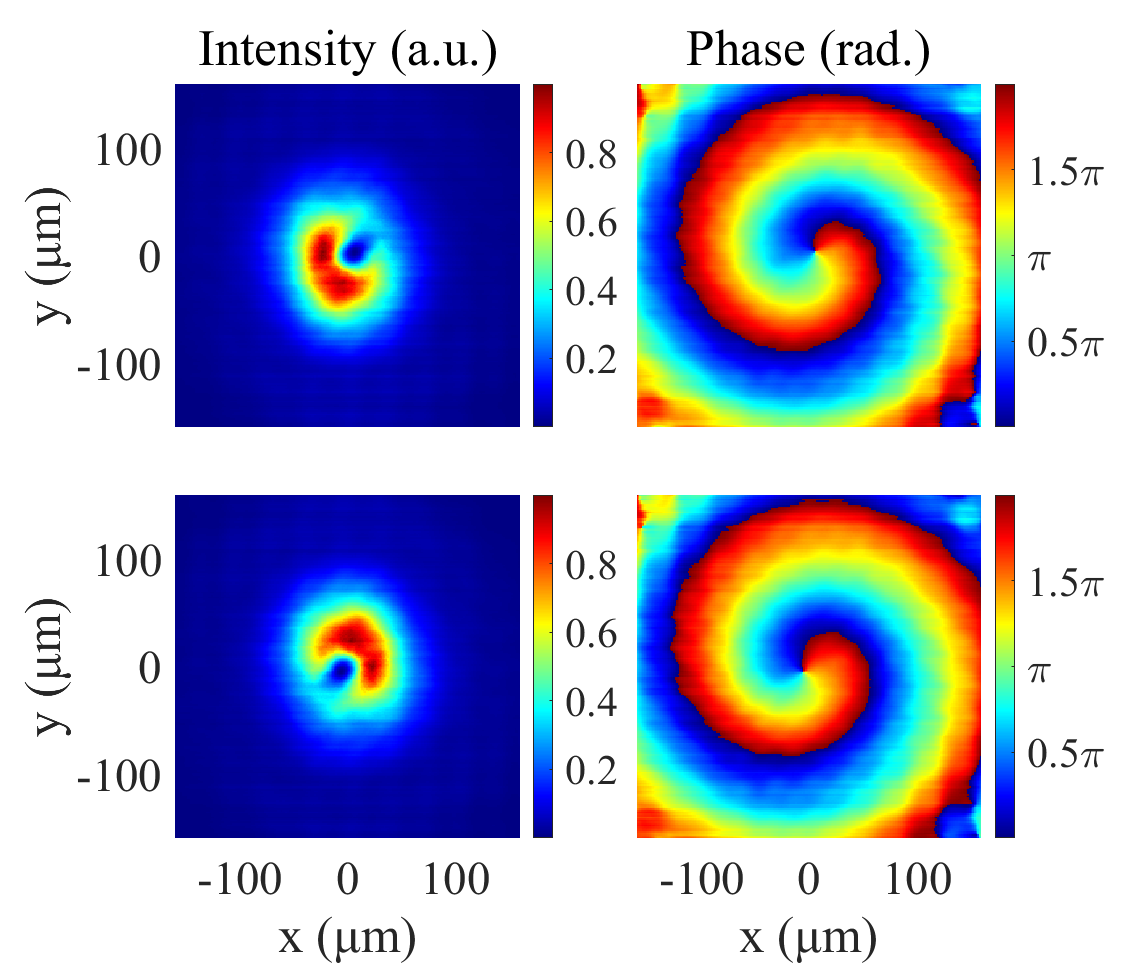}
    \caption{Adjustment precision analysis for the wavefront tilt reflector. Intensity and phase profiles with an angle deviation of -5\% (upper row) and +5\% (lower row).}
    \label{Fig5}
\end{figure}

Other practical factors, such as mirror misalignment and mechanical imperfections in the optical system, as well as drifts in electron beam parameters, can also degrade the quality of the generated vortex beam. To mitigate these issues, it is essential to enhance the robustness and precision of alignment of the optical system. In addition, incorporating real-time feedback systems for both the optical setup and the electron beam can help to maintain optimal conditions. Similar approaches have been successfully implemented in self-seeding experiments, providing valuable insights for addressing such challenges in future experimental validations of the proposed method.

In conclusion, we have proposed a novel method for generating FEL vortices with variable wavelengths. By slightly adjusting the wavefront of the radiation pulse during the FEL gain process, the OAM mode can dominate and be amplified to saturation. This method is particularly well-suited for FEL facilities equipped with self-seeding setups, which naturally incorporate the necessary chicane and optical transport systems required by this approach. For applications in the hard x-ray regime, it may also be possible to employ crystal transmission or reflection to achieve the required wavefront tilt. The proposed method can also be applied in the seed laser system or in the radiator of an external seeded FEL to control the transverse profile of the coherent high-harmonic radiation. Currently, this method is limited to generating vortex beams with topological charges of 1 and -1. For higher topological charges, due to the larger divergence angle of OAM and its weaker coupling with the electron beam, its gain is difficult to surpass that of the Gaussian mode \cite{PhysRevLett.109.224801}. However, we are actively exploring ways to amplify the vortex beam with higher topological charge by disturbing the gain of the Gaussian mode through a multiple-stage wavefront-tilting mirror setup. The proposed method is compatible with not only various operational modes of FELs but also promises to deliver x-ray vortex beams of unprecedented intensity and quality, paving the way for groundbreaking research in x-ray sciences.

The authors thank E. Hemsing, P. R. Ribi\ifmmode \check{c}\else \v{c}\fi{}, J. Yan, H. Sun, K. Zhang, Y. Liu and Z. Zhao for helpful discussions. This work was supported by the National Natural Science Foundation of China (12435011), CAS Project for Young Scientists in Basic Research (YSBR-115), and Strategic Priority Research Program of the CAS (XDB0530000).‌

\bibliographystyle{apsrev4-2}
\bibliography{reference}

\begin{thebibliography}{52}%
\makeatletter
\providecommand \@ifxundefined [1]{%
 \@ifx{#1\undefined}
}%
\providecommand \@ifnum [1]{%
 \ifnum #1\expandafter \@firstoftwo
 \else \expandafter \@secondoftwo
 \fi
}%
\providecommand \@ifx [1]{%
 \ifx #1\expandafter \@firstoftwo
 \else \expandafter \@secondoftwo
 \fi
}%
\providecommand \natexlab [1]{#1}%
\providecommand \enquote  [1]{``#1''}%
\providecommand \bibnamefont  [1]{#1}%
\providecommand \bibfnamefont [1]{#1}%
\providecommand \citenamefont [1]{#1}%
\providecommand \href@noop [0]{\@secondoftwo}%
\providecommand \href [0]{\begingroup \@sanitize@url \@href}%
\providecommand \@href[1]{\@@startlink{#1}\@@href}%
\providecommand \@@href[1]{\endgroup#1\@@endlink}%
\providecommand \@sanitize@url [0]{\catcode `\\12\catcode `\$12\catcode `\&12\catcode `\#12\catcode `\^12\catcode `\_12\catcode `\%12\relax}%
\providecommand \@@startlink[1]{}%
\providecommand \@@endlink[0]{}%
\providecommand \url  [0]{\begingroup\@sanitize@url \@url }%
\providecommand \@url [1]{\endgroup\@href {#1}{\urlprefix }}%
\providecommand \urlprefix  [0]{URL }%
\providecommand \Eprint [0]{\href }%
\providecommand \doibase [0]{https://doi.org/}%
\providecommand \selectlanguage [0]{\@gobble}%
\providecommand \bibinfo  [0]{\@secondoftwo}%
\providecommand \bibfield  [0]{\@secondoftwo}%
\providecommand \translation [1]{[#1]}%
\providecommand \BibitemOpen [0]{}%
\providecommand \bibitemStop [0]{}%
\providecommand \bibitemNoStop [0]{.\EOS\space}%
\providecommand \EOS [0]{\spacefactor3000\relax}%
\providecommand \BibitemShut  [1]{\csname bibitem#1\endcsname}%
\let\auto@bib@innerbib\@empty
%</preamble>
\bibitem [{\citenamefont {Allen}\ \emph {et~al.}(1992)\citenamefont {Allen}, \citenamefont {Beijersbergen}, \citenamefont {Spreeuw},\ and\ \citenamefont {Woerdman}}]{PhysRevA.45.8185}%
  \BibitemOpen
  \bibfield  {author} {\bibinfo {author} {\bibfnamefont {L.}~\bibnamefont {Allen}}, \bibinfo {author} {\bibfnamefont {M.~W.}\ \bibnamefont {Beijersbergen}}, \bibinfo {author} {\bibfnamefont {R.~J.~C.}\ \bibnamefont {Spreeuw}},\ and\ \bibinfo {author} {\bibfnamefont {J.~P.}\ \bibnamefont {Woerdman}},\ }\href {https://doi.org/10.1103/PhysRevA.45.8185} {\bibfield  {journal} {\bibinfo  {journal} {Phys. Rev. A}\ }\textbf {\bibinfo {volume} {45}},\ \bibinfo {pages} {8185} (\bibinfo {year} {1992})}\BibitemShut {NoStop}%
\bibitem [{\citenamefont {Shen}\ \emph {et~al.}(2019)\citenamefont {Shen}, \citenamefont {Wang}, \citenamefont {Xie}, \citenamefont {Min}, \citenamefont {Fu}, \citenamefont {Liu}, \citenamefont {Gong},\ and\ \citenamefont {Yuan}}]{RN1}%
  \BibitemOpen
  \bibfield  {author} {\bibinfo {author} {\bibfnamefont {Y.}~\bibnamefont {Shen}}, \bibinfo {author} {\bibfnamefont {X.}~\bibnamefont {Wang}}, \bibinfo {author} {\bibfnamefont {Z.}~\bibnamefont {Xie}}, \bibinfo {author} {\bibfnamefont {C.}~\bibnamefont {Min}}, \bibinfo {author} {\bibfnamefont {X.}~\bibnamefont {Fu}}, \bibinfo {author} {\bibfnamefont {Q.}~\bibnamefont {Liu}}, \bibinfo {author} {\bibfnamefont {M.}~\bibnamefont {Gong}},\ and\ \bibinfo {author} {\bibfnamefont {X.}~\bibnamefont {Yuan}},\ }\href {https://doi.org/10.1038/s41377-019-0194-2} {\bibfield  {journal} {\bibinfo  {journal} {Light: Science \& Applications}\ }\textbf {\bibinfo {volume} {8}},\ \bibinfo {pages} {90} (\bibinfo {year} {2019})}\BibitemShut {NoStop}%
\bibitem [{\citenamefont {Paterson}\ \emph {et~al.}(2001)\citenamefont {Paterson}, \citenamefont {MacDonald}, \citenamefont {Arlt}, \citenamefont {Sibbett}, \citenamefont {Bryant},\ and\ \citenamefont {Dholakia}}]{doi:10.1126/science.1058591}%
  \BibitemOpen
  \bibfield  {author} {\bibinfo {author} {\bibfnamefont {L.}~\bibnamefont {Paterson}}, \bibinfo {author} {\bibfnamefont {M.~P.}\ \bibnamefont {MacDonald}}, \bibinfo {author} {\bibfnamefont {J.}~\bibnamefont {Arlt}}, \bibinfo {author} {\bibfnamefont {W.}~\bibnamefont {Sibbett}}, \bibinfo {author} {\bibfnamefont {P.~E.}\ \bibnamefont {Bryant}},\ and\ \bibinfo {author} {\bibfnamefont {K.}~\bibnamefont {Dholakia}},\ }\href {https://doi.org/10.1126/science.1058591} {\bibfield  {journal} {\bibinfo  {journal} {Science}\ }\textbf {\bibinfo {volume} {292}},\ \bibinfo {pages} {912} (\bibinfo {year} {2001})}\BibitemShut {NoStop}%
\bibitem [{\citenamefont {MacDonald}\ \emph {et~al.}(2002)\citenamefont {MacDonald}, \citenamefont {Paterson}, \citenamefont {Volke-Sepulveda}, \citenamefont {Arlt}, \citenamefont {Sibbett},\ and\ \citenamefont {Dholakia}}]{doi:10.1126/science.1069571}%
  \BibitemOpen
  \bibfield  {author} {\bibinfo {author} {\bibfnamefont {M.~P.}\ \bibnamefont {MacDonald}}, \bibinfo {author} {\bibfnamefont {L.}~\bibnamefont {Paterson}}, \bibinfo {author} {\bibfnamefont {K.}~\bibnamefont {Volke-Sepulveda}}, \bibinfo {author} {\bibfnamefont {J.}~\bibnamefont {Arlt}}, \bibinfo {author} {\bibfnamefont {W.}~\bibnamefont {Sibbett}},\ and\ \bibinfo {author} {\bibfnamefont {K.}~\bibnamefont {Dholakia}},\ }\href {https://doi.org/10.1126/science.1069571} {\bibfield  {journal} {\bibinfo  {journal} {Science}\ }\textbf {\bibinfo {volume} {296}},\ \bibinfo {pages} {1101} (\bibinfo {year} {2002})}\BibitemShut {NoStop}%
\bibitem [{\citenamefont {Grier}(2003)}]{RN2}%
  \BibitemOpen
  \bibfield  {author} {\bibinfo {author} {\bibfnamefont {D.~G.}\ \bibnamefont {Grier}},\ }\href {https://doi.org/10.1038/nature01935} {\bibfield  {journal} {\bibinfo  {journal} {Nature}\ }\textbf {\bibinfo {volume} {424}},\ \bibinfo {pages} {810} (\bibinfo {year} {2003})}\BibitemShut {NoStop}%
\bibitem [{\citenamefont {Shen}\ \emph {et~al.}(2012)\citenamefont {Shen}, \citenamefont {Hu}, \citenamefont {Yuan}, \citenamefont {Min}, \citenamefont {Fang},\ and\ \citenamefont {Yuan}}]{Shen:12}%
  \BibitemOpen
  \bibfield  {author} {\bibinfo {author} {\bibfnamefont {Z.}~\bibnamefont {Shen}}, \bibinfo {author} {\bibfnamefont {Z.~J.}\ \bibnamefont {Hu}}, \bibinfo {author} {\bibfnamefont {G.~H.}\ \bibnamefont {Yuan}}, \bibinfo {author} {\bibfnamefont {C.~J.}\ \bibnamefont {Min}}, \bibinfo {author} {\bibfnamefont {H.}~\bibnamefont {Fang}},\ and\ \bibinfo {author} {\bibfnamefont {X.-C.}\ \bibnamefont {Yuan}},\ }\href {https://doi.org/10.1364/OL.37.004627} {\bibfield  {journal} {\bibinfo  {journal} {Opt. Lett.}\ }\textbf {\bibinfo {volume} {37}},\ \bibinfo {pages} {4627} (\bibinfo {year} {2012})}\BibitemShut {NoStop}%
\bibitem [{\citenamefont {Erhard}\ \emph {et~al.}(2018)\citenamefont {Erhard}, \citenamefont {Fickler}, \citenamefont {Krenn},\ and\ \citenamefont {Zeilinger}}]{RN3}%
  \BibitemOpen
  \bibfield  {author} {\bibinfo {author} {\bibfnamefont {M.}~\bibnamefont {Erhard}}, \bibinfo {author} {\bibfnamefont {R.}~\bibnamefont {Fickler}}, \bibinfo {author} {\bibfnamefont {M.}~\bibnamefont {Krenn}},\ and\ \bibinfo {author} {\bibfnamefont {A.}~\bibnamefont {Zeilinger}},\ }\href {https://doi.org/10.1038/lsa.2017.146} {\bibfield  {journal} {\bibinfo  {journal} {Light: Science \& Applications}\ }\textbf {\bibinfo {volume} {7}},\ \bibinfo {pages} {17146} (\bibinfo {year} {2018})}\BibitemShut {NoStop}%
\bibitem [{\citenamefont {F\"{u}rhapter}\ \emph {et~al.}(2005)\citenamefont {F\"{u}rhapter}, \citenamefont {Jesacher}, \citenamefont {Bernet},\ and\ \citenamefont {Ritsch-Marte}}]{Furhapter:05}%
  \BibitemOpen
  \bibfield  {author} {\bibinfo {author} {\bibfnamefont {S.}~\bibnamefont {F\"{u}rhapter}}, \bibinfo {author} {\bibfnamefont {A.}~\bibnamefont {Jesacher}}, \bibinfo {author} {\bibfnamefont {S.}~\bibnamefont {Bernet}},\ and\ \bibinfo {author} {\bibfnamefont {M.}~\bibnamefont {Ritsch-Marte}},\ }\href {https://doi.org/10.1364/OPEX.13.000689} {\bibfield  {journal} {\bibinfo  {journal} {Opt. Express}\ }\textbf {\bibinfo {volume} {13}},\ \bibinfo {pages} {689} (\bibinfo {year} {2005})}\BibitemShut {NoStop}%
\bibitem [{\citenamefont {Tamburini}\ \emph {et~al.}(2006)\citenamefont {Tamburini}, \citenamefont {Anzolin}, \citenamefont {Umbriaco}, \citenamefont {Bianchini},\ and\ \citenamefont {Barbieri}}]{PhysRevLett.97.163903}%
  \BibitemOpen
  \bibfield  {author} {\bibinfo {author} {\bibfnamefont {F.}~\bibnamefont {Tamburini}}, \bibinfo {author} {\bibfnamefont {G.}~\bibnamefont {Anzolin}}, \bibinfo {author} {\bibfnamefont {G.}~\bibnamefont {Umbriaco}}, \bibinfo {author} {\bibfnamefont {A.}~\bibnamefont {Bianchini}},\ and\ \bibinfo {author} {\bibfnamefont {C.}~\bibnamefont {Barbieri}},\ }\href {https://doi.org/10.1103/PhysRevLett.97.163903} {\bibfield  {journal} {\bibinfo  {journal} {Phys. Rev. Lett.}\ }\textbf {\bibinfo {volume} {97}},\ \bibinfo {pages} {163903} (\bibinfo {year} {2006})}\BibitemShut {NoStop}%
\bibitem [{\citenamefont {Nagali}\ \emph {et~al.}(2009)\citenamefont {Nagali}, \citenamefont {Sansoni}, \citenamefont {Sciarrino}, \citenamefont {De~Martini}, \citenamefont {Marrucci}, \citenamefont {Piccirillo}, \citenamefont {Karimi},\ and\ \citenamefont {Santamato}}]{RN4}%
  \BibitemOpen
  \bibfield  {author} {\bibinfo {author} {\bibfnamefont {E.}~\bibnamefont {Nagali}}, \bibinfo {author} {\bibfnamefont {L.}~\bibnamefont {Sansoni}}, \bibinfo {author} {\bibfnamefont {F.}~\bibnamefont {Sciarrino}}, \bibinfo {author} {\bibfnamefont {F.}~\bibnamefont {De~Martini}}, \bibinfo {author} {\bibfnamefont {L.}~\bibnamefont {Marrucci}}, \bibinfo {author} {\bibfnamefont {B.}~\bibnamefont {Piccirillo}}, \bibinfo {author} {\bibfnamefont {E.}~\bibnamefont {Karimi}},\ and\ \bibinfo {author} {\bibfnamefont {E.}~\bibnamefont {Santamato}},\ }\href {https://doi.org/10.1038/nphoton.2009.214} {\bibfield  {journal} {\bibinfo  {journal} {Nature Photonics}\ }\textbf {\bibinfo {volume} {3}},\ \bibinfo {pages} {720} (\bibinfo {year} {2009})}\BibitemShut {NoStop}%
\bibitem [{\citenamefont {Mafu}\ \emph {et~al.}(2013)\citenamefont {Mafu}, \citenamefont {Dudley}, \citenamefont {Goyal}, \citenamefont {Giovannini}, \citenamefont {McLaren}, \citenamefont {Padgett}, \citenamefont {Konrad}, \citenamefont {Petruccione}, \citenamefont {L\"utkenhaus},\ and\ \citenamefont {Forbes}}]{PhysRevA.88.032305}%
  \BibitemOpen
  \bibfield  {author} {\bibinfo {author} {\bibfnamefont {M.}~\bibnamefont {Mafu}}, \bibinfo {author} {\bibfnamefont {A.}~\bibnamefont {Dudley}}, \bibinfo {author} {\bibfnamefont {S.}~\bibnamefont {Goyal}}, \bibinfo {author} {\bibfnamefont {D.}~\bibnamefont {Giovannini}}, \bibinfo {author} {\bibfnamefont {M.}~\bibnamefont {McLaren}}, \bibinfo {author} {\bibfnamefont {M.~J.}\ \bibnamefont {Padgett}}, \bibinfo {author} {\bibfnamefont {T.}~\bibnamefont {Konrad}}, \bibinfo {author} {\bibfnamefont {F.}~\bibnamefont {Petruccione}}, \bibinfo {author} {\bibfnamefont {N.}~\bibnamefont {L\"utkenhaus}},\ and\ \bibinfo {author} {\bibfnamefont {A.}~\bibnamefont {Forbes}},\ }\href {https://doi.org/10.1103/PhysRevA.88.032305} {\bibfield  {journal} {\bibinfo  {journal} {Phys. Rev. A}\ }\textbf {\bibinfo {volume} {88}},\ \bibinfo {pages} {032305} (\bibinfo {year} {2013})}\BibitemShut {NoStop}%
\bibitem [{\citenamefont {Firth}\ and\ \citenamefont {Skryabin}(1997)}]{PhysRevLett.79.2450}%
  \BibitemOpen
  \bibfield  {author} {\bibinfo {author} {\bibfnamefont {W.~J.}\ \bibnamefont {Firth}}\ and\ \bibinfo {author} {\bibfnamefont {D.~V.}\ \bibnamefont {Skryabin}},\ }\href {https://doi.org/10.1103/PhysRevLett.79.2450} {\bibfield  {journal} {\bibinfo  {journal} {Phys. Rev. Lett.}\ }\textbf {\bibinfo {volume} {79}},\ \bibinfo {pages} {2450} (\bibinfo {year} {1997})}\BibitemShut {NoStop}%
\bibitem [{\citenamefont {Zhuang}(2004)}]{doi:10.1126/science.1100603}%
  \BibitemOpen
  \bibfield  {author} {\bibinfo {author} {\bibfnamefont {X.}~\bibnamefont {Zhuang}},\ }\href {https://doi.org/10.1126/science.1100603} {\bibfield  {journal} {\bibinfo  {journal} {Science}\ }\textbf {\bibinfo {volume} {305}},\ \bibinfo {pages} {188} (\bibinfo {year} {2004})}\BibitemShut {NoStop}%
\bibitem [{\citenamefont {Wang}\ and\ \citenamefont {Chan}(2014)}]{RN5}%
  \BibitemOpen
  \bibfield  {author} {\bibinfo {author} {\bibfnamefont {S.~B.}\ \bibnamefont {Wang}}\ and\ \bibinfo {author} {\bibfnamefont {C.~T.}\ \bibnamefont {Chan}},\ }\href {https://doi.org/10.1038/ncomms4307} {\bibfield  {journal} {\bibinfo  {journal} {Nature Communications}\ }\textbf {\bibinfo {volume} {5}},\ \bibinfo {pages} {3307} (\bibinfo {year} {2014})}\BibitemShut {NoStop}%
\bibitem [{\citenamefont {Zhao}\ \emph {et~al.}(2017)\citenamefont {Zhao}, \citenamefont {Askarpour}, \citenamefont {Sun}, \citenamefont {Shi}, \citenamefont {Li},\ and\ \citenamefont {Alù}}]{RN6}%
  \BibitemOpen
  \bibfield  {author} {\bibinfo {author} {\bibfnamefont {Y.}~\bibnamefont {Zhao}}, \bibinfo {author} {\bibfnamefont {A.~N.}\ \bibnamefont {Askarpour}}, \bibinfo {author} {\bibfnamefont {L.}~\bibnamefont {Sun}}, \bibinfo {author} {\bibfnamefont {J.}~\bibnamefont {Shi}}, \bibinfo {author} {\bibfnamefont {X.}~\bibnamefont {Li}},\ and\ \bibinfo {author} {\bibfnamefont {A.}~\bibnamefont {Alù}},\ }\href {https://doi.org/10.1038/ncomms14180} {\bibfield  {journal} {\bibinfo  {journal} {Nature Communications}\ }\textbf {\bibinfo {volume} {8}},\ \bibinfo {pages} {14180} (\bibinfo {year} {2017})}\BibitemShut {NoStop}%
\bibitem [{\citenamefont {Takahashi}\ \emph {et~al.}(2013)\citenamefont {Takahashi}, \citenamefont {Suzuki}, \citenamefont {Furutaku}, \citenamefont {Yamauchi}, \citenamefont {Kohmura},\ and\ \citenamefont {Ishikawa}}]{PhysRevB.87.121201}%
  \BibitemOpen
  \bibfield  {author} {\bibinfo {author} {\bibfnamefont {Y.}~\bibnamefont {Takahashi}}, \bibinfo {author} {\bibfnamefont {A.}~\bibnamefont {Suzuki}}, \bibinfo {author} {\bibfnamefont {S.}~\bibnamefont {Furutaku}}, \bibinfo {author} {\bibfnamefont {K.}~\bibnamefont {Yamauchi}}, \bibinfo {author} {\bibfnamefont {Y.}~\bibnamefont {Kohmura}},\ and\ \bibinfo {author} {\bibfnamefont {T.}~\bibnamefont {Ishikawa}},\ }\href {https://doi.org/10.1103/PhysRevB.87.121201} {\bibfield  {journal} {\bibinfo  {journal} {Phys. Rev. B}\ }\textbf {\bibinfo {volume} {87}},\ \bibinfo {pages} {121201} (\bibinfo {year} {2013})}\BibitemShut {NoStop}%
\bibitem [{\citenamefont {Hernández-García}\ \emph {et~al.}(2017)\citenamefont {Hernández-García}, \citenamefont {Vieira}, \citenamefont {Mendonça}, \citenamefont {Rego}, \citenamefont {San~Román}, \citenamefont {Plaja}, \citenamefont {Ribic}, \citenamefont {Gauthier},\ and\ \citenamefont {Picón}}]{photonics4020028}%
  \BibitemOpen
  \bibfield  {author} {\bibinfo {author} {\bibfnamefont {C.}~\bibnamefont {Hernández-García}}, \bibinfo {author} {\bibfnamefont {J.}~\bibnamefont {Vieira}}, \bibinfo {author} {\bibfnamefont {J.~T.}\ \bibnamefont {Mendonça}}, \bibinfo {author} {\bibfnamefont {L.}~\bibnamefont {Rego}}, \bibinfo {author} {\bibfnamefont {J.}~\bibnamefont {San~Román}}, \bibinfo {author} {\bibfnamefont {L.}~\bibnamefont {Plaja}}, \bibinfo {author} {\bibfnamefont {P.~R.}\ \bibnamefont {Ribic}}, \bibinfo {author} {\bibfnamefont {D.}~\bibnamefont {Gauthier}},\ and\ \bibinfo {author} {\bibfnamefont {A.}~\bibnamefont {Picón}},\ }\bibfield  {journal} {\bibinfo  {journal} {Photonics}\ }\textbf {\bibinfo {volume} {4}},\ \href {https://doi.org/10.3390/photonics4020028} {10.3390/photonics4020028} (\bibinfo {year} {2017})\BibitemShut {NoStop}%
\bibitem [{\citenamefont {De~Ninno}\ \emph {et~al.}(2020)\citenamefont {De~Ninno} \emph {et~al.}}]{np-photoelectric}%
  \BibitemOpen
  \bibfield  {author} {\bibinfo {author} {\bibfnamefont {G.}~\bibnamefont {De~Ninno}} \emph {et~al.},\ }\href {https://doi.org/10.1038/s41566-020-0669-y} {\bibfield  {journal} {\bibinfo  {journal} {Nature Photonics}\ }\textbf {\bibinfo {volume} {14}},\ \bibinfo {pages} {554} (\bibinfo {year} {2020})}\BibitemShut {NoStop}%
\bibitem [{\citenamefont {Wang}\ \emph {et~al.}(2023)\citenamefont {Wang}, \citenamefont {Brooks}, \citenamefont {Johnsen}, \citenamefont {Jenkins}, \citenamefont {Esashi}, \citenamefont {Binnie}, \citenamefont {Tanksalvala}, \citenamefont {Kapteyn},\ and\ \citenamefont {Murnane}}]{Wang:23}%
  \BibitemOpen
  \bibfield  {author} {\bibinfo {author} {\bibfnamefont {B.}~\bibnamefont {Wang}}, \bibinfo {author} {\bibfnamefont {N.~J.}\ \bibnamefont {Brooks}}, \bibinfo {author} {\bibfnamefont {P.}~\bibnamefont {Johnsen}}, \bibinfo {author} {\bibfnamefont {N.~W.}\ \bibnamefont {Jenkins}}, \bibinfo {author} {\bibfnamefont {Y.}~\bibnamefont {Esashi}}, \bibinfo {author} {\bibfnamefont {I.}~\bibnamefont {Binnie}}, \bibinfo {author} {\bibfnamefont {M.}~\bibnamefont {Tanksalvala}}, \bibinfo {author} {\bibfnamefont {H.~C.}\ \bibnamefont {Kapteyn}},\ and\ \bibinfo {author} {\bibfnamefont {M.~M.}\ \bibnamefont {Murnane}},\ }\href {https://doi.org/10.1364/OPTICA.498619} {\bibfield  {journal} {\bibinfo  {journal} {Optica}\ }\textbf {\bibinfo {volume} {10}},\ \bibinfo {pages} {1245} (\bibinfo {year} {2023})}\BibitemShut {NoStop}%
\bibitem [{\citenamefont {McCarter}\ \emph {et~al.}(2024)\citenamefont {McCarter}, \citenamefont {Long}, \citenamefont {Hastings},\ and\ \citenamefont {Roy}}]{McCarter_2024}%
  \BibitemOpen
  \bibfield  {author} {\bibinfo {author} {\bibfnamefont {M.~R.}\ \bibnamefont {McCarter}}, \bibinfo {author} {\bibfnamefont {L.~E.~D.}\ \bibnamefont {Long}}, \bibinfo {author} {\bibfnamefont {J.~T.}\ \bibnamefont {Hastings}},\ and\ \bibinfo {author} {\bibfnamefont {S.}~\bibnamefont {Roy}},\ }\href {https://doi.org/10.1088/1361-648X/ad53b3} {\bibfield  {journal} {\bibinfo  {journal} {Journal of Physics: Condensed Matter}\ }\textbf {\bibinfo {volume} {36}},\ \bibinfo {pages} {423003} (\bibinfo {year} {2024})}\BibitemShut {NoStop}%
\bibitem [{\citenamefont {Ivanov}(2022)}]{IVANOV2022103987}%
  \BibitemOpen
  \bibfield  {author} {\bibinfo {author} {\bibfnamefont {I.~P.}\ \bibnamefont {Ivanov}},\ }\href {https://doi.org/https://doi.org/10.1016/j.ppnp.2022.103987} {\bibfield  {journal} {\bibinfo  {journal} {Progress in Particle and Nuclear Physics}\ }\textbf {\bibinfo {volume} {127}},\ \bibinfo {pages} {103987} (\bibinfo {year} {2022})}\BibitemShut {NoStop}%
\bibitem [{\citenamefont {Ackermann}\ \emph {et~al.}(2007)\citenamefont {Ackermann} \emph {et~al.}}]{RN8}%
  \BibitemOpen
  \bibfield  {author} {\bibinfo {author} {\bibfnamefont {W.}~\bibnamefont {Ackermann}} \emph {et~al.},\ }\href {https://doi.org/10.1038/nphoton.2007.76} {\bibfield  {journal} {\bibinfo  {journal} {Nature Photonics}\ }\textbf {\bibinfo {volume} {1}},\ \bibinfo {pages} {336} (\bibinfo {year} {2007})}\BibitemShut {NoStop}%
\bibitem [{\citenamefont {Emma}\ \emph {et~al.}(2010)\citenamefont {Emma} \emph {et~al.}}]{RN9}%
  \BibitemOpen
  \bibfield  {author} {\bibinfo {author} {\bibfnamefont {P.}~\bibnamefont {Emma}} \emph {et~al.},\ }\href {https://doi.org/10.1038/nphoton.2010.176} {\bibfield  {journal} {\bibinfo  {journal} {Nature Photonics}\ }\textbf {\bibinfo {volume} {4}},\ \bibinfo {pages} {641} (\bibinfo {year} {2010})}\BibitemShut {NoStop}%
\bibitem [{\citenamefont {Ishikawa}\ \emph {et~al.}(2012)\citenamefont {Ishikawa} \emph {et~al.}}]{RN7}%
  \BibitemOpen
  \bibfield  {author} {\bibinfo {author} {\bibfnamefont {T.}~\bibnamefont {Ishikawa}} \emph {et~al.},\ }\href {https://doi.org/10.1038/nphoton.2012.141} {\bibfield  {journal} {\bibinfo  {journal} {Nature Photonics}\ }\textbf {\bibinfo {volume} {6}},\ \bibinfo {pages} {540} (\bibinfo {year} {2012})}\BibitemShut {NoStop}%
\bibitem [{\citenamefont {Allaria}\ \emph {et~al.}(2015)\citenamefont {Allaria} \emph {et~al.}}]{article}%
  \BibitemOpen
  \bibfield  {author} {\bibinfo {author} {\bibfnamefont {E.}~\bibnamefont {Allaria}} \emph {et~al.},\ }\href {https://doi.org/10.1107/S1600577515005366} {\bibfield  {journal} {\bibinfo  {journal} {Journal of Synchrotron Radiation}\ }\textbf {\bibinfo {volume} {22}},\ \bibinfo {pages} {485} (\bibinfo {year} {2015})}\BibitemShut {NoStop}%
\bibitem [{\citenamefont {Kang}\ \emph {et~al.}(2017)\citenamefont {Kang}, \citenamefont {Min}, \citenamefont {Heo}, \citenamefont {Kim}, \citenamefont {Yang}, \citenamefont {Kim}, \citenamefont {Nam}, \citenamefont {Baek}, \citenamefont {Choi}, \citenamefont {Mun} \emph {et~al.}}]{RN10}%
  \BibitemOpen
  \bibfield  {author} {\bibinfo {author} {\bibfnamefont {H.-S.}\ \bibnamefont {Kang}}, \bibinfo {author} {\bibfnamefont {C.-K.}\ \bibnamefont {Min}}, \bibinfo {author} {\bibfnamefont {H.}~\bibnamefont {Heo}}, \bibinfo {author} {\bibfnamefont {C.}~\bibnamefont {Kim}}, \bibinfo {author} {\bibfnamefont {H.}~\bibnamefont {Yang}}, \bibinfo {author} {\bibfnamefont {G.}~\bibnamefont {Kim}}, \bibinfo {author} {\bibfnamefont {I.}~\bibnamefont {Nam}}, \bibinfo {author} {\bibfnamefont {S.~Y.}\ \bibnamefont {Baek}}, \bibinfo {author} {\bibfnamefont {H.-J.}\ \bibnamefont {Choi}}, \bibinfo {author} {\bibfnamefont {G.}~\bibnamefont {Mun}}, \emph {et~al.},\ }\href {https://doi.org/10.1038/s41566-017-0029-8} {\bibfield  {journal} {\bibinfo  {journal} {Nature Photonics}\ }\textbf {\bibinfo {volume} {11}},\ \bibinfo {pages} {708} (\bibinfo {year} {2017})}\BibitemShut {NoStop}%
\bibitem [{\citenamefont {Decking}\ \emph {et~al.}(2020)\citenamefont {Decking} \emph {et~al.}}]{RN12}%
  \BibitemOpen
  \bibfield  {author} {\bibinfo {author} {\bibfnamefont {W.}~\bibnamefont {Decking}} \emph {et~al.},\ }\href {https://doi.org/10.1038/s41566-020-0607-z} {\bibfield  {journal} {\bibinfo  {journal} {Nature Photonics}\ }\textbf {\bibinfo {volume} {14}},\ \bibinfo {pages} {391} (\bibinfo {year} {2020})}\BibitemShut {NoStop}%
\bibitem [{\citenamefont {Prat}\ \emph {et~al.}(2020)\citenamefont {Prat} \emph {et~al.}}]{RN13}%
  \BibitemOpen
  \bibfield  {author} {\bibinfo {author} {\bibfnamefont {E.}~\bibnamefont {Prat}} \emph {et~al.},\ }\href {https://doi.org/10.1038/s41566-020-00712-8} {\bibfield  {journal} {\bibinfo  {journal} {Nature Photonics}\ }\textbf {\bibinfo {volume} {14}},\ \bibinfo {pages} {748} (\bibinfo {year} {2020})}\BibitemShut {NoStop}%
\bibitem [{\citenamefont {Feng}\ \emph {et~al.}(2022)\citenamefont {Feng}, \citenamefont {Liu}, \citenamefont {Chen}, \citenamefont {Zhou}, \citenamefont {Zhang}, \citenamefont {Qi}, \citenamefont {Gu}, \citenamefont {Wang}, \citenamefont {Jiang}, \citenamefont {Li} \emph {et~al.}}]{Feng:22}%
  \BibitemOpen
  \bibfield  {author} {\bibinfo {author} {\bibfnamefont {C.}~\bibnamefont {Feng}}, \bibinfo {author} {\bibfnamefont {T.}~\bibnamefont {Liu}}, \bibinfo {author} {\bibfnamefont {S.}~\bibnamefont {Chen}}, \bibinfo {author} {\bibfnamefont {K.}~\bibnamefont {Zhou}}, \bibinfo {author} {\bibfnamefont {K.}~\bibnamefont {Zhang}}, \bibinfo {author} {\bibfnamefont {Z.}~\bibnamefont {Qi}}, \bibinfo {author} {\bibfnamefont {D.}~\bibnamefont {Gu}}, \bibinfo {author} {\bibfnamefont {Z.}~\bibnamefont {Wang}}, \bibinfo {author} {\bibfnamefont {Z.}~\bibnamefont {Jiang}}, \bibinfo {author} {\bibfnamefont {X.}~\bibnamefont {Li}}, \emph {et~al.},\ }\href {https://doi.org/10.1364/OPTICA.466064} {\bibfield  {journal} {\bibinfo  {journal} {Optica}\ }\textbf {\bibinfo {volume} {9}},\ \bibinfo {pages} {785} (\bibinfo {year} {2022})}\BibitemShut {NoStop}%
\bibitem [{\citenamefont {Maroju}\ \emph {et~al.}(2020)\citenamefont {Maroju}, \citenamefont {Grazioli}, \citenamefont {Di~Fraia}, \citenamefont {Moioli}, \citenamefont {Ertel}, \citenamefont {Ahmadi}, \citenamefont {Plekan}, \citenamefont {Finetti}, \citenamefont {Allaria}, \citenamefont {Giannessi} \emph {et~al.}}]{RN19}%
  \BibitemOpen
  \bibfield  {author} {\bibinfo {author} {\bibfnamefont {P.~K.}\ \bibnamefont {Maroju}}, \bibinfo {author} {\bibfnamefont {C.}~\bibnamefont {Grazioli}}, \bibinfo {author} {\bibfnamefont {M.}~\bibnamefont {Di~Fraia}}, \bibinfo {author} {\bibfnamefont {M.}~\bibnamefont {Moioli}}, \bibinfo {author} {\bibfnamefont {D.}~\bibnamefont {Ertel}}, \bibinfo {author} {\bibfnamefont {H.}~\bibnamefont {Ahmadi}}, \bibinfo {author} {\bibfnamefont {O.}~\bibnamefont {Plekan}}, \bibinfo {author} {\bibfnamefont {P.}~\bibnamefont {Finetti}}, \bibinfo {author} {\bibfnamefont {E.}~\bibnamefont {Allaria}}, \bibinfo {author} {\bibfnamefont {L.}~\bibnamefont {Giannessi}}, \emph {et~al.},\ }\href {https://doi.org/10.1038/s41586-020-2005-6} {\bibfield  {journal} {\bibinfo  {journal} {Nature}\ }\textbf {\bibinfo {volume} {578}},\ \bibinfo {pages} {386} (\bibinfo {year} {2020})}\BibitemShut {NoStop}%
\bibitem [{\citenamefont {Duris}\ \emph {et~al.}(2020)\citenamefont {Duris}, \citenamefont {Li}, \citenamefont {Driver}, \citenamefont {Champenois}, \citenamefont {MacArthur}, \citenamefont {Lutman}, \citenamefont {Zhang}, \citenamefont {Rosenberger}, \citenamefont {Aldrich}, \citenamefont {Coffee} \emph {et~al.}}]{RN20}%
  \BibitemOpen
  \bibfield  {author} {\bibinfo {author} {\bibfnamefont {J.}~\bibnamefont {Duris}}, \bibinfo {author} {\bibfnamefont {S.}~\bibnamefont {Li}}, \bibinfo {author} {\bibfnamefont {T.}~\bibnamefont {Driver}}, \bibinfo {author} {\bibfnamefont {E.~G.}\ \bibnamefont {Champenois}}, \bibinfo {author} {\bibfnamefont {J.~P.}\ \bibnamefont {MacArthur}}, \bibinfo {author} {\bibfnamefont {A.~A.}\ \bibnamefont {Lutman}}, \bibinfo {author} {\bibfnamefont {Z.}~\bibnamefont {Zhang}}, \bibinfo {author} {\bibfnamefont {P.}~\bibnamefont {Rosenberger}}, \bibinfo {author} {\bibfnamefont {J.~W.}\ \bibnamefont {Aldrich}}, \bibinfo {author} {\bibfnamefont {R.}~\bibnamefont {Coffee}}, \emph {et~al.},\ }\href {https://doi.org/10.1038/s41566-019-0549-5} {\bibfield  {journal} {\bibinfo  {journal} {Nature Photonics}\ }\textbf {\bibinfo {volume} {14}},\ \bibinfo {pages} {30} (\bibinfo {year} {2020})}\BibitemShut {NoStop}%
\bibitem [{\citenamefont {Hemsing}\ \emph {et~al.}(2009)\citenamefont {Hemsing}, \citenamefont {Musumeci}, \citenamefont {Reiche}, \citenamefont {Tikhoplav}, \citenamefont {Marinelli}, \citenamefont {Rosenzweig},\ and\ \citenamefont {Gover}}]{PhysRevLett.102.174801}%
  \BibitemOpen
  \bibfield  {author} {\bibinfo {author} {\bibfnamefont {E.}~\bibnamefont {Hemsing}}, \bibinfo {author} {\bibfnamefont {P.}~\bibnamefont {Musumeci}}, \bibinfo {author} {\bibfnamefont {S.}~\bibnamefont {Reiche}}, \bibinfo {author} {\bibfnamefont {R.}~\bibnamefont {Tikhoplav}}, \bibinfo {author} {\bibfnamefont {A.}~\bibnamefont {Marinelli}}, \bibinfo {author} {\bibfnamefont {J.~B.}\ \bibnamefont {Rosenzweig}},\ and\ \bibinfo {author} {\bibfnamefont {A.}~\bibnamefont {Gover}},\ }\href {https://doi.org/10.1103/PhysRevLett.102.174801} {\bibfield  {journal} {\bibinfo  {journal} {Phys. Rev. Lett.}\ }\textbf {\bibinfo {volume} {102}},\ \bibinfo {pages} {174801} (\bibinfo {year} {2009})}\BibitemShut {NoStop}%
\bibitem [{\citenamefont {Hemsing}\ \emph {et~al.}(2011)\citenamefont {Hemsing}, \citenamefont {Marinelli},\ and\ \citenamefont {Rosenzweig}}]{PhysRevLett.106.164803}%
  \BibitemOpen
  \bibfield  {author} {\bibinfo {author} {\bibfnamefont {E.}~\bibnamefont {Hemsing}}, \bibinfo {author} {\bibfnamefont {A.}~\bibnamefont {Marinelli}},\ and\ \bibinfo {author} {\bibfnamefont {J.~B.}\ \bibnamefont {Rosenzweig}},\ }\href {https://doi.org/10.1103/PhysRevLett.106.164803} {\bibfield  {journal} {\bibinfo  {journal} {Phys. Rev. Lett.}\ }\textbf {\bibinfo {volume} {106}},\ \bibinfo {pages} {164803} (\bibinfo {year} {2011})}\BibitemShut {NoStop}%
\bibitem [{\citenamefont {Hemsing}\ and\ \citenamefont {Marinelli}(2012)}]{PhysRevLett.109.224801}%
  \BibitemOpen
  \bibfield  {author} {\bibinfo {author} {\bibfnamefont {E.}~\bibnamefont {Hemsing}}\ and\ \bibinfo {author} {\bibfnamefont {A.}~\bibnamefont {Marinelli}},\ }\href {https://doi.org/10.1103/PhysRevLett.109.224801} {\bibfield  {journal} {\bibinfo  {journal} {Phys. Rev. Lett.}\ }\textbf {\bibinfo {volume} {109}},\ \bibinfo {pages} {224801} (\bibinfo {year} {2012})}\BibitemShut {NoStop}%
\bibitem [{\citenamefont {Ribi\ifmmode~\check{c}\else \v{c}\fi{}}\ \emph {et~al.}(2014)\citenamefont {Ribi\ifmmode~\check{c}\else \v{c}\fi{}}, \citenamefont {Gauthier},\ and\ \citenamefont {De~Ninno}}]{PhysRevLett.112.203602}%
  \BibitemOpen
  \bibfield  {author} {\bibinfo {author} {\bibfnamefont {P.~R.}\ \bibnamefont {Ribi\ifmmode~\check{c}\else \v{c}\fi{}}}, \bibinfo {author} {\bibfnamefont {D.}~\bibnamefont {Gauthier}},\ and\ \bibinfo {author} {\bibfnamefont {G.}~\bibnamefont {De~Ninno}},\ }\href {https://doi.org/10.1103/PhysRevLett.112.203602} {\bibfield  {journal} {\bibinfo  {journal} {Phys. Rev. Lett.}\ }\textbf {\bibinfo {volume} {112}},\ \bibinfo {pages} {203602} (\bibinfo {year} {2014})}\BibitemShut {NoStop}%
\bibitem [{\citenamefont {Ribi\ifmmode~\check{c}\else \v{c}\fi{}}\ \emph {et~al.}(2017)\citenamefont {Ribi\ifmmode~\check{c}\else \v{c}\fi{}}, \citenamefont {R\"osner}, \citenamefont {Gauthier}, \citenamefont {Allaria}, \citenamefont {D\"oring}, \citenamefont {Foglia}, \citenamefont {Giannessi}, \citenamefont {Mahne}, \citenamefont {Manfredda}, \citenamefont {Masciovecchio} \emph {et~al.}}]{PhysRevX.7.031036}%
  \BibitemOpen
  \bibfield  {author} {\bibinfo {author} {\bibfnamefont {P.~R.}\ \bibnamefont {Ribi\ifmmode~\check{c}\else \v{c}\fi{}}}, \bibinfo {author} {\bibfnamefont {B.}~\bibnamefont {R\"osner}}, \bibinfo {author} {\bibfnamefont {D.}~\bibnamefont {Gauthier}}, \bibinfo {author} {\bibfnamefont {E.}~\bibnamefont {Allaria}}, \bibinfo {author} {\bibfnamefont {F.}~\bibnamefont {D\"oring}}, \bibinfo {author} {\bibfnamefont {L.}~\bibnamefont {Foglia}}, \bibinfo {author} {\bibfnamefont {L.}~\bibnamefont {Giannessi}}, \bibinfo {author} {\bibfnamefont {N.}~\bibnamefont {Mahne}}, \bibinfo {author} {\bibfnamefont {M.}~\bibnamefont {Manfredda}}, \bibinfo {author} {\bibfnamefont {C.}~\bibnamefont {Masciovecchio}}, \emph {et~al.},\ }\href {https://doi.org/10.1103/PhysRevX.7.031036} {\bibfield  {journal} {\bibinfo  {journal} {Phys. Rev. X}\ }\textbf {\bibinfo {volume} {7}},\ \bibinfo {pages} {031036} (\bibinfo {year} {2017})}\BibitemShut {NoStop}%
\bibitem [{\citenamefont {Sun}\ \emph {et~al.}(2021)\citenamefont {Sun}, \citenamefont {Wang}, \citenamefont {Feng}, \citenamefont {Tu}, \citenamefont {Fan},\ and\ \citenamefont {Liu}}]{Sun_Wang_Feng_Tu_Fan_Liu_2021}%
  \BibitemOpen
  \bibfield  {author} {\bibinfo {author} {\bibfnamefont {H.}~\bibnamefont {Sun}}, \bibinfo {author} {\bibfnamefont {X.}~\bibnamefont {Wang}}, \bibinfo {author} {\bibfnamefont {C.}~\bibnamefont {Feng}}, \bibinfo {author} {\bibfnamefont {L.}~\bibnamefont {Tu}}, \bibinfo {author} {\bibfnamefont {W.}~\bibnamefont {Fan}},\ and\ \bibinfo {author} {\bibfnamefont {B.}~\bibnamefont {Liu}},\ }\href {https://doi.org/10.1017/hpl.2021.52} {\bibfield  {journal} {\bibinfo  {journal} {High Power Laser Science and Engineering}\ }\textbf {\bibinfo {volume} {9}},\ \bibinfo {pages} {e65} (\bibinfo {year} {2021})}\BibitemShut {NoStop}%
\bibitem [{\citenamefont {Yan}\ and\ \citenamefont {Geloni}(2023)}]{yan2023self}%
  \BibitemOpen
  \bibfield  {author} {\bibinfo {author} {\bibfnamefont {J.}~\bibnamefont {Yan}}\ and\ \bibinfo {author} {\bibfnamefont {G.}~\bibnamefont {Geloni}},\ }\href {https://doi.org/10.1117/1.APN.2.3.036001} {\bibfield  {journal} {\bibinfo  {journal} {Advanced Photonics Nexus}\ }\textbf {\bibinfo {volume} {2}},\ \bibinfo {pages} {036001} (\bibinfo {year} {2023})}\BibitemShut {NoStop}%
\bibitem [{\citenamefont {Kondratenko}\ and\ \citenamefont {Saldin}(1980)}]{Kondratenko1980GENERATINGOC}%
  \BibitemOpen
  \bibfield  {author} {\bibinfo {author} {\bibfnamefont {A.~M.}\ \bibnamefont {Kondratenko}}\ and\ \bibinfo {author} {\bibfnamefont {E.~L.}\ \bibnamefont {Saldin}},\ }in\ \href {https://api.semanticscholar.org/CorpusID:122908033} {\emph {\bibinfo {booktitle} {Generating of coherent radiation by a relativistic electron beam in an undulator}}}\ (\bibinfo {year} {1980})\BibitemShut {NoStop}%
\bibitem [{\citenamefont {Bonifacio}\ \emph {et~al.}(1984)\citenamefont {Bonifacio}, \citenamefont {Pellegrini},\ and\ \citenamefont {Narducci}}]{BONIFACIO1984373}%
  \BibitemOpen
  \bibfield  {author} {\bibinfo {author} {\bibfnamefont {R.}~\bibnamefont {Bonifacio}}, \bibinfo {author} {\bibfnamefont {C.}~\bibnamefont {Pellegrini}},\ and\ \bibinfo {author} {\bibfnamefont {L.}~\bibnamefont {Narducci}},\ }\href {https://doi.org/https://doi.org/10.1016/0030-4018(84)90105-6} {\bibfield  {journal} {\bibinfo  {journal} {Optics Communications}\ }\textbf {\bibinfo {volume} {50}},\ \bibinfo {pages} {373} (\bibinfo {year} {1984})}\BibitemShut {NoStop}%
\bibitem [{\citenamefont {Feldhaus}\ \emph {et~al.}(1997)\citenamefont {Feldhaus}, \citenamefont {Saldin}, \citenamefont {Schneider}, \citenamefont {Schneidmiller},\ and\ \citenamefont {Yurkov}}]{soft-x-ray-self-seeding-principle}%
  \BibitemOpen
  \bibfield  {author} {\bibinfo {author} {\bibfnamefont {J.}~\bibnamefont {Feldhaus}}, \bibinfo {author} {\bibfnamefont {E.}~\bibnamefont {Saldin}}, \bibinfo {author} {\bibfnamefont {J.}~\bibnamefont {Schneider}}, \bibinfo {author} {\bibfnamefont {E.}~\bibnamefont {Schneidmiller}},\ and\ \bibinfo {author} {\bibfnamefont {M.}~\bibnamefont {Yurkov}},\ }\href {https://doi.org/10.1016/S0030-4018(97)00163-6} {\bibfield  {journal} {\bibinfo  {journal} {Optics Communications}\ }\textbf {\bibinfo {volume} {140}},\ \bibinfo {pages} {341} (\bibinfo {year} {1997})}\BibitemShut {NoStop}%
\bibitem [{\citenamefont {Ratner}\ \emph {et~al.}(2015)\citenamefont {Ratner}, \citenamefont {Abela}, \citenamefont {Amann} \emph {et~al.}}]{soft-x-ray-self-seeding-experiment}%
  \BibitemOpen
  \bibfield  {author} {\bibinfo {author} {\bibfnamefont {D.}~\bibnamefont {Ratner}}, \bibinfo {author} {\bibfnamefont {R.}~\bibnamefont {Abela}}, \bibinfo {author} {\bibfnamefont {J.}~\bibnamefont {Amann}}, \emph {et~al.},\ }\href {https://doi.org/10.1103/PhysRevLett.114.054801} {\bibfield  {journal} {\bibinfo  {journal} {Phys. Rev. Lett.}\ }\textbf {\bibinfo {volume} {114}},\ \bibinfo {pages} {054801} (\bibinfo {year} {2015})}\BibitemShut {NoStop}%
\bibitem [{\citenamefont {Geloni}\ \emph {et~al.}(2011)\citenamefont {Geloni}, \citenamefont {Kocharyan},\ and\ \citenamefont {Saldin}}]{hard-x-ray-self-seeding-principle}%
  \BibitemOpen
  \bibfield  {author} {\bibinfo {author} {\bibfnamefont {G.}~\bibnamefont {Geloni}}, \bibinfo {author} {\bibfnamefont {V.}~\bibnamefont {Kocharyan}},\ and\ \bibinfo {author} {\bibfnamefont {E.}~\bibnamefont {Saldin}},\ }\href {https://doi.org/10.1080/09500340.2011.586473} {\bibfield  {journal} {\bibinfo  {journal} {Journal of Modern Optics}\ }\textbf {\bibinfo {volume} {58}},\ \bibinfo {pages} {1391–1403} (\bibinfo {year} {2011})}\BibitemShut {NoStop}%
\bibitem [{\citenamefont {Amann}\ \emph {et~al.}(2012)\citenamefont {Amann}, \citenamefont {Berg}, \citenamefont {Blank} \emph {et~al.}}]{hard-x-ray-self-seeding-experiemnt}%
  \BibitemOpen
  \bibfield  {author} {\bibinfo {author} {\bibfnamefont {J.}~\bibnamefont {Amann}}, \bibinfo {author} {\bibfnamefont {W.}~\bibnamefont {Berg}}, \bibinfo {author} {\bibfnamefont {V.}~\bibnamefont {Blank}}, \emph {et~al.},\ }\href {https://doi.org/10.1038/nphoton.2012.180} {\bibfield  {journal} {\bibinfo  {journal} {Nature Photon.}\ }\textbf {\bibinfo {volume} {6}},\ \bibinfo {pages} {693–698} (\bibinfo {year} {2012})}\BibitemShut {NoStop}%
\bibitem [{\citenamefont {Campbell}\ \emph {et~al.}(2012)\citenamefont {Campbell}, \citenamefont {Hage}, \citenamefont {Buchler},\ and\ \citenamefont {Lam}}]{Campbell:12}%
  \BibitemOpen
  \bibfield  {author} {\bibinfo {author} {\bibfnamefont {G.}~\bibnamefont {Campbell}}, \bibinfo {author} {\bibfnamefont {B.}~\bibnamefont {Hage}}, \bibinfo {author} {\bibfnamefont {B.}~\bibnamefont {Buchler}},\ and\ \bibinfo {author} {\bibfnamefont {P.~K.}\ \bibnamefont {Lam}},\ }\href {https://doi.org/10.1364/AO.51.000873} {\bibfield  {journal} {\bibinfo  {journal} {Appl. Opt.}\ }\textbf {\bibinfo {volume} {51}},\ \bibinfo {pages} {873} (\bibinfo {year} {2012})}\BibitemShut {NoStop}%
\bibitem [{\citenamefont {Fickler}\ \emph {et~al.}(2016)\citenamefont {Fickler}, \citenamefont {Campbell}, \citenamefont {Buchler}, \citenamefont {Lam},\ and\ \citenamefont {Zeilinger}}]{10.1073/pnas.1616889113}%
  \BibitemOpen
  \bibfield  {author} {\bibinfo {author} {\bibfnamefont {R.}~\bibnamefont {Fickler}}, \bibinfo {author} {\bibfnamefont {G.}~\bibnamefont {Campbell}}, \bibinfo {author} {\bibfnamefont {B.}~\bibnamefont {Buchler}}, \bibinfo {author} {\bibfnamefont {P.~K.}\ \bibnamefont {Lam}},\ and\ \bibinfo {author} {\bibfnamefont {A.}~\bibnamefont {Zeilinger}},\ }\href {https://doi.org/10.1073/pnas.1616889113} {\bibfield  {journal} {\bibinfo  {journal} {Proceedings of the National Academy of Sciences}\ }\textbf {\bibinfo {volume} {113}},\ \bibinfo {pages} {13642} (\bibinfo {year} {2016})}\BibitemShut {NoStop}%
\bibitem [{\citenamefont {Terhalle}\ \emph {et~al.}(2011)\citenamefont {Terhalle}, \citenamefont {Langner}, \citenamefont {P\"{a}iv\"{a}nranta}, \citenamefont {Guzenko}, \citenamefont {David},\ and\ \citenamefont {Ekinci}}]{Terhalle:11}%
  \BibitemOpen
  \bibfield  {author} {\bibinfo {author} {\bibfnamefont {B.}~\bibnamefont {Terhalle}}, \bibinfo {author} {\bibfnamefont {A.}~\bibnamefont {Langner}}, \bibinfo {author} {\bibfnamefont {B.}~\bibnamefont {P\"{a}iv\"{a}nranta}}, \bibinfo {author} {\bibfnamefont {V.~A.}\ \bibnamefont {Guzenko}}, \bibinfo {author} {\bibfnamefont {C.}~\bibnamefont {David}},\ and\ \bibinfo {author} {\bibfnamefont {Y.}~\bibnamefont {Ekinci}},\ }\href {https://doi.org/10.1364/OL.36.004143} {\bibfield  {journal} {\bibinfo  {journal} {Opt. Lett.}\ }\textbf {\bibinfo {volume} {36}},\ \bibinfo {pages} {4143} (\bibinfo {year} {2011})}\BibitemShut {NoStop}%
\bibitem [{\citenamefont {Goodman}(1969)}]{Goodman1969IntroductionTF}%
  \BibitemOpen
  \bibfield  {author} {\bibinfo {author} {\bibfnamefont {J.~W.}\ \bibnamefont {Goodman}},\ }in\ \href {https://api.semanticscholar.org/CorpusID:118908270} {\emph {\bibinfo {booktitle} {Introduction to Fourier optics}}}\ (\bibinfo  {publisher} {Roberts and Company},\ \bibinfo {year} {1969})\BibitemShut {NoStop}%
\bibitem [{\citenamefont {Serkez}\ \emph {et~al.}(2015)\citenamefont {Serkez}, \citenamefont {Krzywinski}, \citenamefont {Ding},\ and\ \citenamefont {Huang}}]{PhysRevSTAB.18.030708}%
  \BibitemOpen
  \bibfield  {author} {\bibinfo {author} {\bibfnamefont {S.}~\bibnamefont {Serkez}}, \bibinfo {author} {\bibfnamefont {J.}~\bibnamefont {Krzywinski}}, \bibinfo {author} {\bibfnamefont {Y.}~\bibnamefont {Ding}},\ and\ \bibinfo {author} {\bibfnamefont {Z.}~\bibnamefont {Huang}},\ }\href {https://doi.org/10.1103/PhysRevSTAB.18.030708} {\bibfield  {journal} {\bibinfo  {journal} {Phys. Rev. ST Accel. Beams}\ }\textbf {\bibinfo {volume} {18}},\ \bibinfo {pages} {030708} (\bibinfo {year} {2015})}\BibitemShut {NoStop}%
\bibitem [{\citenamefont {Reiche}(1999)}]{REICHE1999243}%
  \BibitemOpen
  \bibfield  {author} {\bibinfo {author} {\bibfnamefont {S.}~\bibnamefont {Reiche}},\ }\href {https://doi.org/https://doi.org/10.1016/S0168-9002(99)00114-X} {\bibfield  {journal} {\bibinfo  {journal} {Nucl. Instrum. Methods Phys. Res. Sec. A: Accel. Spectrom. Detect. Assoc. Equip.}\ }\textbf {\bibinfo {volume} {429}},\ \bibinfo {pages} {243} (\bibinfo {year} {1999})}\BibitemShut {NoStop}%
\bibitem [{\citenamefont {Shu}\ \emph {et~al.}(2019)\citenamefont {Shu}, \citenamefont {Andrade}, \citenamefont {Anton} \emph {et~al.}}]{Mechanical-system-SR}%
  \BibitemOpen
  \bibfield  {author} {\bibinfo {author} {\bibfnamefont {D.}~\bibnamefont {Shu}}, \bibinfo {author} {\bibfnamefont {V.~D.}\ \bibnamefont {Andrade}}, \bibinfo {author} {\bibfnamefont {J.}~\bibnamefont {Anton}}, \emph {et~al.},\ }in\ \href {https://doi.org/10.1117/12.2529384} {\emph {\bibinfo {booktitle} {X-Ray Nanoimaging: Instruments and Methods IV}}},\ \bibinfo {organization} {International Society for Optics and Photonics}\ (\bibinfo  {publisher} {SPIE},\ \bibinfo {year} {2019})\ p.\ \bibinfo {pages} {111120N}\BibitemShut {NoStop}%
\bibitem [{\citenamefont {Shu}\ \emph {et~al.}(2021)\citenamefont {Shu}, \citenamefont {Anton}, \citenamefont {Assoufid} \emph {et~al.}}]{Mechanical-system-CXFEL}%
  \BibitemOpen
  \bibfield  {author} {\bibinfo {author} {\bibfnamefont {D.}~\bibnamefont {Shu}}, \bibinfo {author} {\bibfnamefont {J.}~\bibnamefont {Anton}}, \bibinfo {author} {\bibfnamefont {L.}~\bibnamefont {Assoufid}}, \emph {et~al.},\ }in\ \href {https://doi.org/10.18429/JACoW-MEDSI2020-TUOA02} {\emph {\bibinfo {booktitle} {MEDSI2020}}}\ (\bibinfo  {publisher} {JACoW},\ \bibinfo {year} {2021})\ p.\ \bibinfo {pages} {TUOA02}\BibitemShut {NoStop}%
\end{thebibliography}%

\end{document}